# Gluon Fragmentation into $^3P_J$ Quarkonium

J.P. Ma

Recearch Center for High Energy Physics
School of Physics
University of Melbourne
Parkville, Victoria 3052
Australia

**Abstract**:
The functions of the gluon fragmentation into $^3P_J$ quarkonium are calculated to order $\alpha_s^2$. With the recent progress in analysing quarkonium systems in QCD we show explicitly how the socalled divergence in the limit of the zero-binding energy, which is related to $P$-wave quarkonia, is treated correctly in the case of fragmentation functions. The obtained fragmentation functions satisfy explicitly at the order of $\alpha_s^2$ the Altarelli-Parisi equation and when $z \to 0$ they behave as $z^{-1}$ as expected. Some comments on the previous results are made.

# 1. Introduction

Quantum Chromodynamics(QCD) is believed to be the correct theory for the strong interaction, but its predictive power is limited by the fact that the fundamental quanta, the parton, i.e. the quarks and gluons, are not observed freely in Nature. We observe in the real world only hadrons. This fact makes it impossible to calculate directly from QCD for example an inclusive cross section of single hadron production in a reaction. However, in some kinematical regions, the dominant contribution to the cross section can be decomposed within the framework of QCD into three parts: the partonic part, the part of parton fragmentation into the produced hadron and the part of the parton distributions in the initial hadrons. The partonic part can be calculated perturbatively thanks to the asymptotic freedom of QCD, while all the long distance physics due to the bound state nature of the hadrons is lumped into the parton fragmentation functions and the parton distributions. This is the content of the factorization theorem[see Ref. [1] and references cited therein].

It is hard to study fragmentation functions directly from QCD because of their nonperturbative nature. However, the situation changes if the hadron into which partons fragment is a quarkonium. It is well known that the bound state effect of a quarkonium system can be well described by a nonrelativistic wavefunction if one takes the quarkonium system as a color-singlet bound state of a heavy quark $Q$ and its antiquark $\bar{Q}$, since in the quarkonium rest frame the quark and the antiquark move with a small velocity $v$. Assuming the long distance effect is only contained in the wavefunction, one can calculate fragmentation functions perturbatively in QCD[2–8] except a nonperturbative constant: the wavefunction at the origin or its derivative at the origin, where one makes an expansion in the small parameter $v$. This gives us an opportunity to get an insight into the fragmentation functions by starting directly from QCD, hence such study is important. The importance can also be shown by the fact that the heavy quark fragmentation function for $B_c$ meson obtained in this way has similar behaviour as that from the successful model for the heavy quark fragmentation[9]. But if one calculates the functions of the gluon fragmentation into $P$-wave quarkonia, the result is infrared (I.R.) divergent at the leading order of $\alpha_s$[6,7]. Such I.R. divergence is actually well known in the decay of the $P$-wave quarkonium as the divergence in the limit of the zero-binding energy[10]. The appearance of the I.R. divergence indicates that the wavefunction can not contain all the long distance effect. It should be kept in mind that the calculations with the mentioned assumption are merely calculations within a model, namely the color-singlet model. They are not calculations within the framework of QCD.

In the color-singlet model one only takes for $P$-wave quarkonium production the $P$-wave state of the color singlet $Q\bar{Q}$ into account. The probability for production through this way is proportional to $v^2$. However in [11,12] one realized that a $S$-wave state of a color octet $Q\bar{Q}$ will contribute to the production rates at the same order of $v$ as the $P$-wave state of the color singlet $Q\bar{Q}$, because the probability for generating a $S$-wave state of a color octet $Q\bar{Q}$ is proportional to $v^0$ and this state has a probability at order of $v^2$ to be transmitted into a $P$-wave state of a color singlet $Q\bar{Q}$ through emission of soft gluons. This state can also be combined with a soft gluon to form with a probability at order of $v^2$ one of the components of a quarkonium state. From the point of view of a relativistic quantum field theory a quarkonium state consists not only of a color singlet $Q\bar{Q}$ pair but also of many other components. The $S$-wave state of a color octet $Q\bar{Q}$ combined with a gluon is a possible component. Therefore one should take the color octet $Q\bar{Q}$ into account. Recently, quarkonium systems have been treated rigorously in QCD[12] and new factorization formulas for production and decay rates are obtained. Further, the nonperturbative effect is represented by matrix elements defined in nonrelativistic QCD(NRQCD), not in terms of the wavefunction as in the color-singlet model. For $P$-wave quarkonia there are two nonperturbative constants at the leading order of $v$ instead one in the color-singlet model. The second constant is for the contribution from a color octet $Q\bar{Q}$.

In this work we will calculate with the new factorization formula the functions of the gluon fragmentation into $^3P_J$ quarkonia to order of $\alpha_s^2$, where $J = 0, 1, 2$ is the quantum number of the total angular momentum. We assume that the polarization of quarkonia is not observed and we will work only at the leading order in $v$. We will start from the definition[13] of the fragmentation functions as in [6]. In Sect. 2 we introduce the definition and the new factorization formula for $^3P_J$ quarkonium. The renormalization of the fragmentation functions is briefly discussed. The result for the gluon fragmentation at the leading order of $\alpha_s$ is obtained. For details of the main content in Sect. 2 we refer to Ref[13,12]. In Sect. 3 a perturbative calculation in NRQCD for matrix elements is presented. The results are needed to extract the perturbative parts in the gluon fragmentation and they are universal, i.e. can be used for any process where a $P$-wave quarkonium is produced. In Sect.4 and Sect.5 we extend our calculations to order $\alpha_s^2$. Sect.6 presents our discussion and the summary.

Throughout our work we will use dimensional regularization to regularize ultraviolet(U.V.) divergences and also infrared(I.R.) divergences. With this regularization we maintain the gauge-invariance.

## 2. The Definition of The Gluon Fragmentation Function and The Leading Order Result

We give the definition of the gluon fragmentation function in $d$ dimensions. To give the definition it is convenient to work in the light-cone coordinate system. In this coordinate system a $d$-vector $p$ is expressed as $p^\mu = (p^+, p^-, \mathbf{p_T})$, with $p^+ = (p^0 + p^{d-1})/\sqrt{2}$, $p^- = (p^0 - p^{d-1})/\sqrt{2}$. Introducing a vector $n$ with $n^\mu = (0, 1, 0, \cdots, 0) = (0, 1, \mathbf{0_T})$, the gluon fragmentation function for a spinless hadron $H$ or for a hadron without observing its polarization is defined as[13]:

$$D_{H/G}^{(0)}(z) = \frac{-z^{d-3}}{16(d-2)\pi k^+} \int dx^- e^{-iP^+ x^-/z} < 0|G^{b,+\nu}(0)$$
$$\{\bar{P}\exp\{-ig_s \int_0^\infty d\lambda n \cdot G(\lambda n^\mu)\}\}^{bc} a_H^\dagger(P^+, \mathbf{0_T}) a_H(P^+, \mathbf{0_T}) \quad (2.1)$$
$$\{P\exp\{ig_s \int_{x^-}^\infty d\lambda n \cdot G(\lambda n^\mu)\}\}^{cd} G_\nu^{d,+}(0, x^-, \mathbf{0_T})|0>,$$

where $G_\mu(x) = G_\mu^a(x)T^a$, $G_\mu^a(x)$ is the gluon field and $T^a$ ($a = 1, \ldots, 8$) are the color matrices. $G^{a,\mu\nu}$ is the gluon field strength tensor and $a_H^\dagger(\mathbf{P})$ is the creation operator for the hadron $H$. For hadrons with nonzero spin the summation over the spin is understood. The definition is a unrenormalized version. In calculating $D_{H/G}^{(0)}(z)$ one will encounter U.V. divergences, requiring renormalization. Following [13] the renormalized gluon fragmentation function can be defined as:

$$D_{H/G}(z) = D_{H/G}^{(0)}(z) + \sum_{a=G,q} \int_z^1 \frac{dy}{y} L_a(\frac{y}{z}) D_{H/a}^{(0)}(y). \quad (2.2)$$

Here the function $L_a(z)$ is chosen so as to cancel the U.V. divergences. In MS scheme $L_a(z)$ takes the form:

$$L_a(z) = \sum_N \frac{1}{\epsilon^N} L_a^{(N)}(z), \quad (2.3)$$

where $\epsilon = 4 - d$. From Eq.(2.2) one can derive the Altarelli-Parisi type of the evolution equation for the fragmentation function. We will use the modified MS scheme. That means in practical calculations that the terms with $N_\epsilon = \frac{2}{\epsilon} - \gamma + \ln(4\pi)$ will be cancelled by the function $L_a(z)$. The function $D_{H/G}(z)$ is interpreted as the probability of a gluon $G$ with momentum $k$ to decay into the hadron $H$ with momentum component $P^+ = zk^+$, it is gauge invariant from the definition. Further the function is also invariant under a Lorentz boost along the moving direction of the hadron and under a rotation with the direction as the rotate axis.

Now we give the new factorization formula for $^3P_J$ quarkonium production. We use the notation $\chi_J$ for the $^3P_J$ quarkonium. A production rate for $\chi_J$ takes the following factorized form[12]:

$$\sigma(\chi_J) = \frac{F_1(^3P_J)}{M^4} <0|O_1^{\chi_J}(^3P_J)|0> + \frac{F_8(^3S_1)}{M^2} <0|O_8^{\chi_J}(^3S_1)|0>. \qquad (2.4)$$

The operators $O_1^{\chi_J}(^3P_J)$ and $O_8^{\chi_J}(^3P_1)$ are given by:

$$\begin{aligned}
O_8^H(^3S_1) &= \chi^\dagger \sigma_i T^a \psi (a_H^\dagger a_H) \psi^\dagger \sigma_i T^a \chi \\
O_1^H(^3P_0) &= \frac{1}{3}\chi^\dagger \left(-\frac{i}{2}\overleftrightarrow{\mathbf{D}}\cdot\sigma\right)\psi (a_H^\dagger a_H) \psi^\dagger \left(-\frac{i}{2}\overleftrightarrow{\mathbf{D}}\cdot\sigma\right)\chi \\
O_1^H(^3P_1) &= \frac{1}{2}\chi^\dagger \left(-\frac{i}{2}\overleftrightarrow{\mathbf{D}}\times\sigma\right)_i\psi (a_H^\dagger a_H) \psi^\dagger \left(-\frac{i}{2}\overleftrightarrow{\mathbf{D}}\times\sigma\right)_i\chi \\
O_1^H(^3P_2) &= \chi^\dagger \left(-\frac{i}{2}\overleftrightarrow{D}_{\{i}\sigma_{j\}}\right)\psi (a_H^\dagger a_H) \psi^\dagger \left(-\frac{i}{2}\overleftrightarrow{D}_{\{i}\sigma_{j\}}\right)\chi,
\end{aligned} \qquad (2.5)$$

where $\mathbf{D}$ is the space part of the covariant derivative $D^\mu$ and $\sigma_i(i=1,2,3)$ is the Pauli matrix. $T^a(a=1,\ldots,8)$ is the $SU(3)$ color matrix. The notation $\{ij\}$ means only to take the symmetric and traceless part of a tensor. In Eq. (2.5) $\psi$ and $\chi^\dagger$ are fields with two components for the quark $Q$ and the antiquark $\bar{Q}$ in NRQCD. $M$ is the mass of the quark. $a_H^+$ is the creation operator for the hadron in its rest frame. The matrix elements in Eq.(2.4) are defined in NRQCD. In Eq.(2.4) the part with $F_8$ is the contribution from the color octet $Q\bar{Q}$ in a $^3S_1$ state. The matrix elements represent the nonperturbative effect, while the coefficients $F_1$ and $F_8$ can be calculated perturbatively and they should be free from I.R. singularities.

For calculating the coefficients $F_1$ and $F_8$ one can takes an on-shell $Q\bar{Q}$ pair with small relative momentum instead of the hadron in Eq. (2.4), and calculates $\sigma(Q\bar{Q})$ with perturbation theory. Then one also calculates the matrix elements in NRQCD with the same $Q\bar{Q}$ pair. Substituting the results into Eq.(2.4) one can identify the coefficients. This is called matching. There will be I.R. divergences in $\sigma(Q\bar{Q})$ and also these I.R. divergences will appear in the matrix elements so that the coefficients are free from them.

According to Eq.(2.4) we can write the form for the gluon fragmentation function as:

$$\begin{aligned}
D_{\chi_J/G}(z) &= \hat{D}_1(z,J)\frac{1}{M^5}<0|O_1^{\chi_J}(^3P_J)|0> + \hat{D}_8(z)\frac{1}{M^3}<0|O_8^{\chi_J}(^3S_1)|0> \\
&= D_{\chi_J/G,1}(z) + D_{\chi_J/G,8}(z).
\end{aligned} \qquad (2.6)$$

The function $\hat{D}_1(z,J)$ and $\hat{D}_8(z)$ are dimensionless. In Eq.(2.6) the contribution from a color octet $Q\bar{Q}$ pair and from a color singlet $Q\bar{Q}$ pair is given by $D_{\chi_J/G,8}(z)$ and $D_{\chi_J/G,1}(z)$ respectively, with corresponding perturbative parts $\hat{D}_8(z)$ and $\hat{D}_1(z, J)$. We may call them the color-octet and color-singlet components. The function $\hat{D}_8(z)$ is same for $J = 0, 1, 2$. The task of the present work is to determine $\hat{D}_8(z)$ and $\hat{D}_1(z, J)$ to order of $\alpha_s^2$.

At order of $\alpha_s$, only $\hat{D}_8(z)$ is nonzero, while $\hat{D}_1(z)$ becomes nonzero at order of $\alpha_s^2$. At leading order of $\alpha_s$ there is one Feynman diagram in Fig.1 which contributes to $D_{\chi_J/G,8}(z)$. In Fig. 1 the double line presents the line operator in Eq.(2.1), the broken line is the Cutkosky cut. The calculation can be easily done with the matching procedure mentioned before. We obtain

$$\hat{D}_8(z) = \frac{\pi}{24}\alpha_s(\mu)\delta(1-z) \tag{2.7}$$

This result is independent of $d$. To calculate contributions in the next order of $\alpha_s$, we need to calculate the matrix element $<0|O_8^{Q\bar{Q}}(^3S_1)|0>$ to order of $\alpha_s$, where the hadron is replaced by a $Q\bar{Q}$ pair as required by the matching. We will do the calculation in the next section.

## 3. The Matrix Element in NRQCD

At the tree-level, the contribution to the matrix element $<0|O_8^{Q\bar{Q}}(^3S_1)|0>$ is presented by the diagram in Fiq.2a, where the black circle means the vertex sandwiched between $\chi$ and $\psi^\dagger$ in the operator. For the matrix element here it is $\sigma_i T^a$. The line means the quark line. The broken line is the Cutkosky cut. We let the $Q\bar{Q}$ pair on the left side of the diagram move with a relative velocity $2\mathbf{v}$ and the $Q\bar{Q}$ pair in the right side of the diagram move with a relative velocity $2\mathbf{v}'$. The $Q\bar{Q}$ pair is in CMS. At the tree-level we obtain

$$<0|O_8^{Q\bar{Q}}(^3S_1)|0> = \eta^\dagger \sigma_i T^a \xi \xi^\dagger \sigma_i T^a \eta, \tag{3.1}$$

where $\eta^\dagger$ and $\xi$ are the Pauli spinors with two components in NRQCD for the quark and antiquark respectively.

In the next order there are two types of the diagrams contributing to the matrix elements. One type is given in Fig.2b. This type of the diagram includes the correction to the vertex and the correction to the external quark line. The other type is given in Fig.2c, where a real gluon is emitted by the quarks. This causes the transition from a color octet $Q\bar{Q}$ into a color singlet $Q\bar{Q}$.

Using NRQCD we calculate the contribution from these two types of the diagrams. In the calculation we make an expansion in $v = |\mathbf{v}|$ and $v' = |\mathbf{v}'|$ and neglect in the expansion irrelevant higher order terms. The contribution from the type of the diagrams in Fig.2b before the renormalization is:

$$M_{Fig.2b} = \frac{g_s^2}{(2\pi)^2} \eta^\dagger \sigma_i T^a \xi \xi^\dagger \sigma_i T^a \eta \cdot \{(3 - \frac{4}{9}(v^2 + v'^2))(\frac{1}{\epsilon_I} + \frac{1}{\epsilon} + \ln(\frac{\mu}{\mu_I})) \\ - \frac{\pi^2}{24v}(1 + v^2) - \frac{\pi^2}{24v'}(1 + v'^2)\} \quad (3.2)$$

where $\frac{1}{\epsilon}$ is the U.V. divergence and $\mu$ is the corresponding scale, $\frac{1}{\epsilon_I}$ means the I.R. divergence with $\epsilon_I = d - 4$ and $\mu_I$ is the corresponding scale. We will keep this notation in our work. The terms with $\frac{1}{v}$ and $\frac{1}{v'}$ are the Coulomb singularities.

The contribution from the type of the diagrams in Fig.2c is:

$$M_{Fig.2c} = \frac{g_s^2}{(2\pi^2)}(\frac{1}{\epsilon_I} + \frac{1}{\epsilon} + \ln(\frac{\mu}{\mu_I})) \\ \cdot \{-(3 + (v^2 + v'^2)\eta^\dagger \sigma_i T^a \xi \xi^\dagger \sigma_i T^a \eta + \frac{4}{3}\mathbf{v} \cdot \mathbf{v}'(\frac{4}{9}\eta^\dagger \sigma_i \xi \xi^\dagger \sigma_i \eta \\ + \frac{5}{6}\eta^\dagger \sigma_i T^a \xi \xi^\dagger \sigma_i T^a \eta)\}. \quad (3.3)$$

In Eq.(3.3) the terms with $\mathbf{v} \cdot \mathbf{v}'$ in the second line can be identified as the matrix elements of the operators $O_1^{Q\bar{Q}}(^3P_J)$. These terms present the transition mentioned before. For renormalization one introduces for the operators a matrix of the renormalization constants because of the operator mixing. After renormalization the result is free from U.V. divergence. The I.R. divergence still remains and it represents the nonperturbative nature of the matrix element. We obtain:

$$< 0|O_8^{Q\bar{Q}}(^3S_1)|0 > = \left\{1 - \frac{g_s^2}{(2\pi)^2}(\frac{\pi^2}{24v} + \frac{\pi^2}{24v'})\right\} < 0|O_8^{Q\bar{Q}}(^3S_1)|0 >_0 \\ + \frac{g_s^2}{(2\pi)^2} \cdot \frac{16}{27} \cdot \frac{1}{M^2} < 0|O_1^{Q\bar{Q}}(^3P_0) + O_1^{Q\bar{Q}}(^3P_1) + O_1^{Q\bar{Q}}(^3P_2)|0 >_0 \\ \cdot \left\{\frac{1}{\epsilon_I} + \frac{1}{2}(\gamma - \ln(4\pi)) - \ln(\frac{\mu_I}{2M}) + \ln(\frac{\mu}{2M})\right\} + \cdots, \quad (3.4)$$

where the $\cdots$ is for terms in higher orders in $g_s$, in the velocity $v$ and $v'$ and also for a term related to a operator which is irrelevant in this work. The subcript $_0$ with the matrix elements means only to take their tree-level results. A by-product generated in the calculation is the renormalization group equation for the matrix element $< 0|O_8^{\chi_J}(^3S_1)|0 >$. It reads:

$$\mu \frac{\partial < 0|O_8^{\chi_J}(^3S_1)|0 >}{\partial \mu} = \alpha_s(\mu) \frac{16}{27\pi M^2} < 0|O_1^{\chi_J}(^3P_J)|0 > \quad (3.5)$$

This equation was first derived in [12,11].

The result in Eq.(3.4) is useful not only for our calculation in the following sections but also for $^3P_J$ production which may be not due to fragmentation. Using this result we can now calculate higher order contributions to the fragmentation functions in Eq.(2.6). At higher order of $\alpha_s$ it may be difficult to perform the matching procedure mentioned in Sect.2. It is still convenient to use wavefunctions to project out different states from a $Q\bar{Q}$ pair. However, the final results for $\hat{D}_1(z,J)$ and $\hat{D}_8(z)$ do not depend on these wavefunction. Details about such projections may be found in [14]. We introduce a radial wavefunction $R_1(r)$ to project the $^3P_J$ color singlet $Q\bar{Q}$ state and a octet radial wavefunction $R_8^{(a)}(r)(a=1,\ldots,8)$ to project the $^3S_1$ color octet $Q\bar{Q}$ state. With these wavefunctions we obtain for the matrix elements at the tree-level in Eq.(2.5):

$$<0|O_1^{\chi_J}(^3P_J)|0>_0 = \frac{9(2J+1)}{2\pi}|R_1'(0)|^2,$$
$$<0|O_8^{\chi_J}(^3S_1)|0>_0 = \frac{3}{8\pi}\sum_c |R_8^{(c)}(0)|^2, \tag{3.6}$$

where $R_1'(0)$ is the first derivative of $R_1(r)$ at the origin. The result in Eq.(3.4) can be expressed with these wavefunctions as:

$$<0|O_8^{\chi_J}(^3S_1)|0> = \frac{3}{8\pi}\sum_c |R_8^{(c)}(0)|^2 (1-\frac{g_s^2}{(2\pi)^2}\frac{\pi^2}{12}<v^{-1}>)$$
$$+\frac{g_s^2}{(2\pi)^2}\cdot\frac{16}{27}\cdot\frac{1}{M^2}\frac{9(2J+1)}{2\pi}|R_1'(0)|^2 \tag{3.7}$$
$$\{\frac{1}{\epsilon_I}+\frac{1}{2}(\gamma-\ln(4\pi)-\ln(\frac{\mu_I}{2M})+\ln(\frac{\mu}{2M})\}.$$

The quantity $<v^{-1}>$ is the average of $v^{-1}$ with the color octet wavefunction.

## 4. The Color-Singlet Component

In this section we explain our calculation in detail for $J=1$ and simply give the results for $J=0,2$. As mentioned in Sect. 3 we will use the color singlet wavefunction to project out the $^3P_1$ color singlet $Q\bar{Q}$ state. In order to present the Feynman diagrams in the calculation efficiently, we decompose the function $D_{H/G}(z)$ as:

$$D_{H/G}(z) \sim T_H \otimes T_H^\dagger. \tag{4.1}$$

Here the symbol $\otimes$ denotes contractions of color- and Lorentz-indices and summations over all possible intermediate states. Such a decomposition is always possible by sandwiching the operator $\sum_X |X><X|$ between $a_H^\dagger$ and $a_H$ in the definition in Eq.(2.1). $T_H$ may be called fragmentation amplitude.

The color-singlet component $D_{\chi_J/G,1}(z)$ becomes nonzero at the order of $\alpha_s^2$. The Feynman diagrams for $T_H$ are given in Fig.3. Since the $Q\bar{Q}$ pair is in a color singlet state, there must at least be two gluon lines attaching the quark line. There is a I.R. divergence if the energy of the gluon exchanged between two quarks becomes zero. This results a singularity in $D_{\chi_J/G,1}(z)$ at $z=1$. We use dimensional regularization to regularize this I.R. divergence. The calculation is tedious but straightforward. We obtain:

$$D_{\chi_J/G,1}(z) = \left(\frac{27}{2\pi M^5}|R_1'(0)|^2\right)\frac{1}{81} \cdot \frac{g_s^4}{8\pi^2}$$
$$\cdot \left\{\left(\frac{1}{\epsilon_I} + \frac{1}{2}(\gamma - \ln(4\pi))\right) - \ln(\frac{\mu_I}{2M}) + \frac{5}{8}\right)\delta(1-z) + \frac{1}{(1-z)_+} - 1 - \frac{1}{4}z - z^2\right\}. \quad (4.2)$$

Here the I.R. singularity comes from the term $(1-z)^{-1+\epsilon_I}$. We used the expansion in $\epsilon_I$ for the distribution $(1-z)^{-1+\epsilon_I}$

$$(1-z)^{-1+\epsilon_I} = \frac{1}{\epsilon_I}\delta(1-z) + \frac{1}{(1-z)_+} + \epsilon_I\left(\frac{\ln(1-z)}{(1-z)}\right)_+ + \cdots, \quad (4.3)$$

and neglected termes in higher order in $\epsilon_I$. The + prescription is the same as in [15]. To extract the perturbative part $\hat{D}_1(z, J=1)$ in Eq. (2.6) one notes that there is a contribution from the color singlet wavefunction in $<0|O_8^{\chi_1}(^3S_1)|0>$ at the one-loop level in Eq.(3.7) and we obtain at the order of $\alpha_s^2$ or $g_s^4$:

$$D_{\chi_1/G,1}(z) = \hat{D}_1(z, J=1)\frac{27}{2\pi}|R_1'(0)|^2\frac{1}{M^5}$$
$$+ \frac{\pi}{24}\alpha_s(\mu) \cdot \frac{g_s^2}{(2\pi)^2} \cdot \frac{16}{27} \cdot \frac{1}{M^5}\frac{27}{2\pi}|R_1'(0)|^2\delta(1-z) \quad (4.4)$$
$$\left\{\frac{1}{\epsilon_I} + \frac{1}{2}(\gamma - \ln(4\pi) - \ln(\frac{\mu_I}{2M}) + \ln(\frac{\mu}{2M})\right\}.$$

Comparing Eq.(4.4) and Eq.(4.2) we note the I.R. divergence in the r.h.s. of Eq.(4.4) appears with the same coefficient in the l.h.s., i.e. in Eq.(4.2). The physical meaning here is clear. The $Q\bar{Q}$ pair before emission of the real gluon is in a $S$-wave color-octet state. When the gluon becomes soft, the color octet $Q\bar{Q}$ pair becomes real and the emission of the soft gluon is nonperturbative. This nonperturbative effect is correctly represented by the matrix element $<0|O_8^{\chi_1}(^3S_1)|0>$, and hence $\hat{D}_1(z,1)$ is I.R. finite and it is:

$$\hat{D}_1(z,1) = \frac{2}{81}\alpha_s^2(\mu)\left\{[\frac{5}{8} - \ln(\frac{\mu}{2M})]\delta(1-z) + \frac{1}{(1-z)_+} - 1 - \frac{1}{4}z - z^2\right\}. \quad (4.5)$$

Performing the cacluation for $J = 0, 2$ we also obtain $\hat{D}_1(z, J = 0, 2)$, which are finite,

$$\begin{aligned}
\hat{D}_1(z, J = 0) &= \frac{2}{81}\alpha_s^2(\mu)\Big\{\big[\frac{1}{4} - \ln(\frac{\mu}{2M})\big]\delta(1-z) + \frac{1}{(1-z)_+} \\
&\quad + \frac{9}{4}(5 - 3z)\ln(1-z) + \frac{1}{8}(-26z^2 + 85z - 8)\Big\} \\
\hat{D}_1(z, J = 2) &= \frac{2}{81}\alpha_s^2(\mu)\Big\{\big[\frac{19}{40} - \ln(\frac{\mu}{2M})\big]\delta(1-z) + \frac{1}{(1-z)_+} \\
&\quad + \frac{9}{5}(2 - z)\ln(1-z) + \frac{1}{4}(-4z^2 + 11z - 4)\Big\}.
\end{aligned} \tag{4.6}$$

The results in Eq.(4.5) and Eq.(4.7) are actually not new. They were first obtained in [6] for $J = 1$ and in [7] for all $J$. However, the results here do not agree with those in [6,7] at $z = 1$. For $z \neq 1$ they agree. The disagreement clearly depends on how the I.R. divergence is regularized and some comments are in order. In [6] a cutoff for the transversal momentum of the exchanged gluon is introduced for the regularization and the singularity is absorbed by the matrix element through a renormalization group equation. As we can see from the calculation in Sect.3, such cutoff is meanless since the transversal direction can not be defined and the introduction of the cutoff violates the rotation invariance. Therefore the result in [6] is not correct for $z = 1$. In [7] a cutoff $\Lambda$ of the gluon energy is introduced, and this cutoff is regarded as a scale for separating short distance effects at the energy scale M from long distance effects at the scale $Mv$ or smaller in quarkonium systems. Hence in the result from [7] there is a term for the I.R. singularity like $\sim \delta(1-z)\ln\Lambda$. Such term is actually cancelled if one takes for the matrix element the one-loop result, where the same cutoff should be used. The effect may be only to replace $\Lambda$ with the renormalization scale $\mu$. However such a cutoff can violate the gauge invariance and the invariance of a Lorentz boost in the hadron moving direction. The disagreement may be due to these violations. We used the dimensional regularization and the gauge invariance is respected in each step. Generally, extracted perturbative results are different by separating short distance effects from long distance effects through introducing explicitly a separating scale, which usually is some cutoff, or through identifying I.R. divergences as long distance effects in dimensional regularization, since the definition of nonperturbative parts, which although is hard to give, is different. An example may be found in theoretical attempts to explain EMC effect[16]. In [7] the coefficient for the term with $\delta(1-z)$ is $\frac{13}{12} - \ln\frac{\Lambda}{M}$, $\frac{23}{24} - \ln\frac{\Lambda}{M}$ and $\frac{121}{120} - \ln\frac{\Lambda}{M}$ for $J = 0, 1$ and 2 respectively. Comparing these with our corresponding results $\frac{1}{4} - \ln\frac{\mu}{2M}$, $\frac{5}{8} - \ln\frac{\mu}{2M}$ and $\frac{19}{40} - \ln\frac{\mu}{2M}$ one can see that the disagreement is different for different $J$. This means that one can not relate the cutoff $\Lambda$ to $\mu$ with a simple relation to solve the discrepancy and the discrepancy can not be absorbed into the definition of the color-octet matrix element.

## 5. The Color-Octet Component

In this section we proceed to calculate the color-component $D_{\chi J/G,8}(z)$, or $\hat{D}_8(z)$ upto order of $\alpha_s^2$. This component is independent of the total angular momentum of the quarkonium. We use the color octet wavefunction to project out the color octet $Q\bar{Q}$ state. According to the structure of the Feynman diagrams we write $D_{\chi J/G,8}(z)$ with the result in Sect. 2 as the following form:

$$D_{\chi J/G,8}(z) = \frac{g_s^2}{96}\delta(1-z)\frac{3}{8\pi}\sum_c |R_8^c(0)|^2 \frac{1}{M^3} \\ + D_A(z) + D_B(z) + D_C(z). \tag{5.1}$$

Here we neglected terms of order higher than $\alpha_s^2$. The term in the first line is the result from Sect. 2. The function $D_A(z)$, $D_B(z)$ and $D_C(z)$ will now be calculated

$D_A(z)$ is the contribution by adding the conventional one loop correction for the quark-gluon vertex, the gluon propagator and the external quark line in Fig.1. The U.V. divergences are simply cancelled by the renormalization constants of the coupling and the wavefunctions. We obtain

$$D_A(z) = \frac{g_s^4}{(4\pi)^2}\frac{3}{8\pi}\sum_c |R_8^c(0)|^2 \frac{1}{M^3}\frac{1}{16} \cdot \left\{ \left(\frac{8}{3} - \frac{2}{9}N_f\right)\ln\left(\frac{\mu^2}{4M^2}\right) \right.\\ -\frac{\pi^2}{18}<v^{-1}> + \left(\frac{2}{\epsilon_I} + \gamma - \ln(4\pi) - \ln\left(\frac{\mu_I^2}{4M^2}\right)\right) \\ + 1.822 + \frac{4}{3}\sum_q^{N_f} \text{Re} F_P\left(\frac{m_q^2}{4M^2}\right) \right\}\delta(1-z), \tag{5.2}$$

$$F_P(x) = \int_0^1 d\alpha(\alpha - \alpha^2)\ln\left(\alpha^2 - \alpha + x - i0^+\right)$$

where $N_f$ is the total number of the quark flavours in QCD. The Coulomb singularity $<v^{-1}>$ comes only from the vertex correction. The term with the function $F_P$ comes from the correction of the gluon propagator through quark loops. This function will have an imaginary part if $x < 1$ and only the real part contributes. The number 1.822.. comes from some factors like $\ln 2$ and $\ln 3$. In $D_A(z)$ there is a I.R. singularity.

$D_B(z)$ is the contribution from the Feynman diagrams given in Fig.4a for the fragmentation amplitude $T_H$. This contribution is proportional to $\delta(1-z)$. In this part there is a U.V. divergence and also I.R. divergences as the poles $\epsilon^{-2}$ and $\epsilon^{-1}$. Here some details in the calculation may be needed to give. In the calculation we have such type of the integral over a loop momentum $q$

$$\int \left(\frac{dq}{2\pi}\right)^d \frac{1}{(P+q)^2+i0^+} \cdot \frac{1}{q^2+i0^+} \cdot \frac{1}{-n\cdot q+i0^+} \left\{ f_0(n\cdot q) + \frac{f_1(n\cdot q)}{P\cdot q + q^2 + i0^+} \right\} \quad (5.3)$$

where $P$ has only $+$ and $-$ components and $P^2 = 4M^2$. The term with $f_0$ is U.V. divergent, whereas both terms are I.R. divergent. To calculate the integral one can perform first the integration over $q^-$ via Cauchy's theorem and then integrate over $\mathbf{q_T}$. The U.V. divergence will appear after the $\mathbf{q_T}$ integration as $\Gamma(\frac{\epsilon}{2})$. In order to identify various divergences we will keep the factor $(d-2)$ and factors like $\Gamma$-function combined with $\pi$, which are generated by the $\mathbf{q_T}$ integration. The remained factors are expanded in $\epsilon$. The result for the contribution before renormalization is:

$$D_B^{(0)}(z) = -\frac{g_s^4}{(4\pi)^2} \frac{3}{8\pi} \sum_c |R_8^{(c)}(0)|^2 \frac{1}{M^3} \frac{1}{16} \frac{2}{d-2} (4\pi)^{\frac{4-d}{2}} \left(\frac{\mu}{2M}\right)^{4-d}$$

$$\cdot \left\{ \left(\frac{4}{\epsilon^2} - \frac{2}{\epsilon} - 4\ln 2 - \pi^2\right) \cdot \Gamma(1+\frac{\epsilon}{2}) \right. \quad (5.4)$$

$$\left. + \left(-\frac{4}{\epsilon} + 1 + \epsilon(\frac{1}{2} + \frac{\pi^2}{6})\right) \cdot \Gamma(\frac{\epsilon}{2}) \right\} \delta(1-z).$$

Here we neglected irrelevant terms in higher order of $\epsilon$. The singularity in $\Gamma(\frac{\epsilon}{2})$ is the U.V. one. Other singularities are I.R. singularities.

Now we turn to the last part $D_C(z)$. This part receives contributions from the diagrams in Fig.3 and also from those in Fig.4b. In this part there is a real gluon exchange and we have to integrate over its transversal momentum. The integration over the $+$ component and $-$ component of the gluon momentum is easily done by the $\delta$-functions, one of which is the on-shell condition and the other is due to the total momentum conservation in the $+$ direction. Again there are U.V. divergence and I.R. divergences. We obtain:

$$D_C^{(0)}(z) = \frac{g_s^4}{(4\pi)^2} \frac{3}{8\pi} \sum_c |R_8^{(c)}(0)|^2 \frac{1}{M^3} \frac{1}{16} \frac{2}{d-2} (4\pi)^{\frac{4-d}{2}} \left(\frac{\mu}{2M}\right)^{4-d}$$

$$\cdot \left\{ \left(\frac{4}{\epsilon^2}\delta(1-z) - \delta(1-z) + \left(\frac{\ln(1-z)}{1-z}\right)_+ (2+z-2z^2+z^3)\right) \cdot \Gamma(1+\frac{\epsilon}{2}) \right.$$

$$+ \left(-\frac{4}{\epsilon} + 2\delta(1-z) + \frac{2}{(1-z)_+} z^{-1}(1-2z+3z^2-2z^3+z^4)\right) \quad (5.5)$$

$$\left. - \epsilon\left[\frac{1}{(1-z)_+} + \left(\frac{\ln(1-z)}{1-z}\right)_+\right] z^{-1}(1-2z+3z^2-2z^3+z^4)\right) \cdot \Gamma(\frac{\epsilon}{2}) \right\}.$$

The U.V. divergence is $\Gamma(\frac{\epsilon}{2})$. Other poles in $\epsilon$ are I.R. singularities. Adding $D_C^{(0)}(z)$ and $D_B^{(0)}(z)$ together the I.R. singularity with $\epsilon^{-2}$ is compensated. Further, the I.R.

singularity in the coefficients in the front of $\Gamma(\frac{\epsilon}{2})$ is cancelled too. Now we can safely perform renormalization according to Eq. (2.2) with the result we obtain

$$\begin{aligned}D_B(z) + D_C(z) =& \frac{g_s^4}{16(4\pi)^2}\frac{3}{8\pi}\sum_c |R_8^{(c)}(0)|^2 \frac{1}{M^3} \cdot \{[\delta(1-z) \\
&+ \frac{2}{(1-z)_+}z^{-1}(1-2z+3z^2-2z^3+z^4)]\ln\Big(\frac{\mu^2}{4M^2}\Big) \\
&+ \big[-\frac{2}{\epsilon_I} + \ln 4\pi - \gamma + \ln\Big(\frac{\mu_I^2}{4M^2}\Big)\big]\delta(1-z) \\
&+ \big(4\ln 2 + \frac{2\pi^2}{3}\big)\delta(1-z) \\
&- z^{-1}(2-4z+z^2-z^3)\ln(1-z)\}\end{aligned}$$ (5.6)

where we still have a I.R. singularity $\epsilon_I^{-1}$. However this singularity will be cancelled if we add $D_A(z)$. The final result for $D_{\chi J/G,8}(z)$ is:

$$\begin{aligned}D_{\chi J/G,8}(z) =& \frac{g_s^2}{96}\delta(1-z)\frac{3}{8\pi}\sum_c |R_8^{(c)}(0)|^2 \frac{1}{M^3} \\
&+ \frac{g_s^4}{16(4\pi)^2}\frac{3}{8\pi}\sum_c |R_8^{(c)}(0)|^2 \frac{1}{M^3} \cdot \{[(\frac{11}{3} - \frac{2}{9}N_f)\delta(1-z) \\
&+ \frac{2}{(1-z)_+}z^{-1}(1-2z+3z^2-2z^3+z^4)]\ln\Big(\frac{\mu^2}{4M^2}\Big) \\
&- z^{-1}(2-4z+z^2-z^3)\ln(1-z) \\
&+ \big[-\frac{\pi^2}{18}<v^{-1}> + 4.4946 + \frac{2\pi^2}{3} + \frac{4}{3}\sum_q^{N_f}\mathrm{Re}F_P\Big(\frac{m_q^2}{4M^2}\Big)\big]\delta(1-z)\}.\end{aligned}$$ (5.7)

To extract $\hat{D}_8(z)$ we use the result of the quark matrix element in Sect. 2 and obtain the equation for $\hat{D}_8(z)$ to order $\alpha_s^2$

$$\begin{aligned}D_{\chi J/G,8}(z) =& \frac{g_s^2}{96}\delta(1-z)\frac{3}{8\pi}\sum_c |R_8^{(c)}(0)|^2 \frac{1}{M^3}(1 - \frac{g_s^2}{(2\pi)^2}\frac{\pi^2}{12}<v^{-1}>) \\
&+ (\hat{D}_8(z) - \frac{g_s^2}{96}\delta(1-z))\frac{3}{8\pi}\sum_c |R_8^{(c)}(0)|^2 \frac{1}{M^3}.\end{aligned}$$ (5.8)

We note that the Coulomb singularity in Eq.(5.8) appears with the same coefficient as in Eq.(5.7). This fact results that $\hat{D}_8(z)$ is free from this singularity. From Eq.(5.8) and (5.7)

one finds that

$$\hat{D}_8(z) = \frac{\pi}{24}\alpha_s(\mu)\delta(1-z)$$
$$+ \frac{\alpha_s^2(\mu)}{16} \cdot \{[(\frac{11}{3} - \frac{2}{9}N_f)\delta(1-z)$$
$$+ \frac{2}{(1-z)_+}z^{-1}(1-2z+3z^2-2z^3+z^4)]\ln(\frac{\mu^2}{4M^2}) \quad (5.9)$$
$$- z^{-1}(2-4z+z^2-z^3)\ln(1-z)$$
$$+ [4.5946 + \frac{2\pi^2}{3} + \frac{4}{3}\sum_q^{N_f} \text{Re}F_P(\frac{m_q^2}{4M^2})]\delta(1-z)\}.$$

Now we complete our calculation for gluon fragmentation into $^3P_J$-quarkonia. The fragmentation function is

$$D_{\chi_J/G}(z,\mu) = \hat{D}_1(z,J) <0|O_1^{\chi_J}(^3P_J)|0> \frac{1}{M^5} + \hat{D}_8(z) <0|O_8^{\chi_J}(^3S_1)|0> \frac{1}{M^3} \quad (5.10)$$

with $\hat{D}_1(z,J)$ and $\hat{D}_8(z)$ given in Eq.(4.5), Eq.(4.6) and Eq.(5.9). Each component is now known to order $\alpha_s^2$. The nonperturbative effect is included in the two matrix elements. It is interesting to check whether the fragmentation functions calculated here satisfy their evolution equation. For gluon fragmentation the evolution equation reads

$$\mu\frac{\partial D_{H/G}(z,\mu)}{\partial \mu} = \int_z^1 \frac{dy}{y} P_{G\to G}(z/y,\mu)D_{H/G}(y,\mu) + \sum_q \int_z^1 \frac{dy}{y} P_{G\to q}(z/y,\mu)D_{H/q}(y,\mu)$$
(5.11)

In one loop approximation the splitting functions $P_{G\to G}(x,\mu)$ and $P_{G\to q}(x,\mu)$ are the same as these in the evolution of parton distributions:

$$P_{G\to G}(x,\mu) = \frac{6\alpha_s(\mu)}{\pi}\left\{\frac{x}{(1-x)_+} + \frac{1}{x} - 1 + x - x^2 + (\frac{11}{12} - \frac{N_f}{18})\delta(1-z)\right\},$$
$$P_{G\to q}(x,\mu) = \frac{\alpha_s(\mu)}{2\pi}\left(x^2 + (1-x)^2\right).$$
(5.12)

Taking the derivative of our gluon fragmentation function we get:

$$\mu\frac{\partial D_{\chi_J/G}}{\partial \mu}(z,\mu) = <0|O_8^{\chi_J}(^3S_1)|0> \frac{1}{M^3}\frac{\alpha_s^2(\mu)}{8} \cdot \{[\frac{11}{6} - \frac{2}{18}N_f]\delta(1-z)$$
$$+ \frac{2}{(1-z)_+}z^{-1}(1-2z+3z^2-2z^3+z^4)\}.$$
(5.13)

The $\mu$-dependence in the color-singlet component $\hat{D}_1(z,J)$ is compensated by that in the matrix element of $O_8^{\chi_J}(^3S_1)$ through the renormalization group equation in Eq.(3.5).

At order of $\alpha_s^2$ the nonzero contribution to the r.h.s. of Eq.(5.11) is only from gluon fragmentation function because quarks can not fragment into the quarkonium at order of $\alpha_s$. Substituting the lowest order result of the gluon fragmentation function into the r.h.s. in Eq.(5.11) we find after a little algebra, that our fragmentation function satisfies Eq.(5.11) at order of $\alpha_s^2$. Here we make some comments on the results for the fragmentation of $c$-quark into $\chi_{cJ}$ given in [8]. Since a gluon can decay into $^3P_J$ quarkonia at order of $\alpha_s$, there must be a $\mu$-dependence in the function of the quark fragmentation into $^3P_J$ quarkonia at order of $\alpha_s^2$ according to the Altarelli-Parisi equation. However, from the results in [8], there is no $\mu$-dependence at this order. Hence the results in [8] for $c$ quark fragmentation into $\chi_{cJ}$ are not correct. A new calculation is needed and is already on the way[17], where it is realized that light quarks can also decay into $^3P_J$ quarkonia through the color octet $Q\bar{Q}$ state at order of $\alpha_s^2$.

From our results we can see that the gluon fragmentation functions behave like $z^{-1}$ as $z \to 0$. This is expected in general or simply from the evolution equation (5.11) and (5.12). However, our results tell us that at the scale $\mu = 2M$ such behaviour disappears.

## 6. Discussion and Summary

To discuss possible impact of our results for $^3P_J$ quarkonium production, we calculate the first moment $M^{(0)}$ of the fragmentation functions. This moment is the probability to find a quarkonium without observing its momentum in a jet initialized by a gluon. A rough estimate for $^3P_J$ quarkonium production due to fragmentation can be made by taking the product of parton cross sections and the first moments of the partons. We take the $c$ quark as an example. We neglect the heavy flavor, i.e. $b$ and $t$ quark and take $N_f = 4$ and $m_q = 0$ for light flavors. To avoid the singularity at $z = 0$, we calculate the moments at $\mu = 2M_c$. We obtain:

$$M^{(0)}(G \to \chi_{c0}) = \frac{1}{M_c^3} <0|O_8^{\chi_{c0}}(^3S_1)|0> \cdot (0.131\alpha_s(2M_c) + 0.537\alpha_s^2(2M_c))$$
$$- 0.067\alpha_s^2(2M_c)\frac{1}{M_c^5}<0|O_1^{\chi_{c0}}(^3P_0)|0>,$$
$$M^{(0)}(G \to \chi_{c1}) = \frac{1}{M_c^3} <0|O_8^{\chi_{c1}}(^3S_1)|0> \cdot (0.131\alpha_s(2M_c) + 0.537\alpha_s^2(2M_c))$$
$$- 0.021\alpha_s^2(2M_c)\frac{1}{M_c^5}<0|O_1^{\chi_{c1}}(^3P_1)|0>,$$

$$M^{(0)}(G \to \chi_{c2}) = \frac{1}{M_c^3} <0|O_8^{\chi_{c2}}(^3S_1)|0> \cdot (0.131\alpha_s(2M_c) + 0.537\alpha_s^2(2M_c))$$
$$- 0.043\alpha_s^2(2M_c)\frac{1}{M_c^5} <0|O_1^{\chi_{c2}}(^3P_1)|0> .$$
(6.1)

It is interesting to note that the color singlet component gives a negative contribution. The matrix elements are until now not known precisely. For different $J$ they are related to each other through a spin factor if one neglects higher order terms in $v$ in NRQCD action. A rough estimation for these matrix elements can be found in [11,12]. Using the results there and $M_c = 1.5$GeV, $\alpha_s(2M_c) = 0.26$, the values of the first moments are $0.9 \times 10^{-4}$, $3.8 \times 10^{-4}$ and $5.5 \times 10^{-4}$ for $\chi_{c0}$, $\chi_{c1}$ and $\chi_{c2}$ respectively. The matrix elements may be possibly determined in lattice study. Their corresponding matrix elements in decays have been studied on lattice in [18]. However it is problematic to study the matrix elements here for production, because in spite of the complicated structure there is a sum of all possible intermediate state combined with a quarkonium. It remains to see how such matrix elements can be studied on lattice.

In this work we have shown how the infrared singularity, which appears in the color singlet model, is correctly treated in the case of the gluon fragmentation functions in the context of recent progress. In the color-singlet model the nonperturbative effect in $^3P_J$ quarkonia is not well separated and some important contribution is missing, thus the infrared singularity appears. In the rigorous treatment in the framework of QCD, such singularity disappears as we have shown. Further, the Coulomb singularity is also absorbed by the matrix elements. The final results we obtained are free from infrared singularities and Coulomb singularities. A clear separation between long distance effect and short distance effect is achieved. In old calculations within the color-singlet model for $^3P_J$ quarkonium production and decay such separation can not be established and the contribution from color octet $Q\bar{Q}$ pairs is neglected. In that sence, new calculations must be done in the way shown here. The results in Sect. 3 may be needed in these new calculations. It is interesting to note a color-octet component can receive contributions in a production at lower order of $\alpha_s$ than a color-singlet component. This fact may be useful for example to explain the discrepancy between theory and experiment for direct $\chi_1$ production at hadron colliders[see [19] and references cited therein], where the important partonic process, i.e., $\chi_1$ production via gluon fusion, proceeds at order of $\alpha_s^3$ in the color-singlet model. If one takes the contribution from color octet $Q\bar{Q}$ pairs, the process can happen at order of $\alpha_s^2$. This maybe explains why the production rate predicted within the color-singlet model is lower than indicated by experimental data. It should be pointed out that the Landau-Yang theorem does not apply here because gluons have color as extra freedom than photons have.

Although full calculations for production rates of quarkonia can in general be performed, but in cases where the factorization theorem applies the amount of work can be greatly reduced by using the concept of fragmentation. One needs only to calculate parton production rates and use fragmentation functions. Nevertheless the importance of our work is not limited only to this issue as a good approximation. As already pointed out in the introduction, with quarkonia one can start directly from QCD to study properties of fragmentation functions and the results can be used to build models for fragmentation into hadrons consisting of one heavy quark with light quarks. The gluon fragmentation functions are generally expected to be singular as $z \to 0$. Through our work here, the obtained fragmentation functions are regular at the whole range of $z$ at the scale $\mu = 2M$.

In the literature there are different methods for calculating fragmentation functions. In [3,5] a method is proposed by separating a fragmentation amplitude and a parton production amplitude in a full amplitude of quarkonium production. The separation is achieved at the tree-level by choosing the light cone gauge: $n \cdot G = 0$[5]. This method is equivalent to the method here. If we choose the light cone gauge, the path-ordered exponential of the gluon field in the definition of a parton fragmentation function is unity, and we can obtain the same Feynman diagrams and the same fragmentation amplitudes presented in [3,5]. In the light cone gauge the number of Feynman diagrams can be reduced considerably. In our case only one diagram in Fig.4b remains in this gauge while the others in Fig.4a and Fig.4b do not exist anymore. But it is not convenient to work with this gauge because the divergences are very difficult to handle. In our work we have taken Feynman gauge.

To summarize: We have calculated to order $\alpha_s^2$ the functions of gluon fragmentation into $^3P_J$ quarkonia, the results are given in Eq.(4.5), Eq.(4.6), Eq.(5.9) and Eq.(5.10). The perturbative parts of the functions are free from infrared and Coulomb singularities. These singularities are correctly factorized into a set of matrix elements, which are defined in NRQCD and represent nonperturbative effect. The obtained fragmentation functions satisfy the Altarelli-Parisi evolution equation explicitly as they must. An useful result for a matrix element in NRQCD is also obtained, which may be needed to calculate cross sections of single $\chi_J$ production.

**Acknowledgment:**

The author would like to thank Prof. O. Nachtmann for useful discussions. Dr. A. Rawlinson is acknowledged for reading the manuscript carefully. This work is supported

## References

[1] J.C. Collins, D.E. Soper and G. Sterman, in Perturbative Quantum Chromodynamics, edited by A.H. Mülller, World Scientific, Singapore, 1989.

[2] Chao-Hsi Chang and Yu-Qi Chen, Phys. Rev. D46 (1992) 3845
C.R. Ji and F. Amiri, Phys. Rev. D35 (1987) 3318

[3] E. Braaten and T.C. Yuan, Phys. Rev. Lett. 71 (1993) 1673

[4] A.F. Falk, M. Luke, M.J. Savage and M.B. Wise, Phys. Lett. B312 (1993) 486

[5] E. Braaten, K. Cheung and T.C. Yuan, Phys. Rev. D48 (1993) 4230, R5049

[6] J.P. Ma, Phys. Lett. B332 (1994) 398

[7] E. Braaten and T.C. Yuan, Phys. Rev. D50 (1994) 3176

[8] T.C. Yuan, Phys. Rev. D50 (1994) 5664

[9] C. Peterson, D. Schlatter, I. Schmidt and P. Zerwas, Phys. Rev. D27 (1983) 105

[10] R. Barbieri, R. Gatto and E. Remiddi, Phys. Lett. B61 (1976) 465
R. Barbieri, M. Caffo and E. Remiddi, Nucl Phys. B162 (1980) 220

[11] G.T. Bodwin, E. Braaten, T.C. Yuan and G.P. Lepage, Phys. Rev. D46 (1992) R3703
G.T. Bodwin, E. Braaten and G.P. Lepage, Phys. Rev. D46 (1992) R1914

[12] G.T. Bodwin, E. Braaten and G.P. Lepage, Phys. Rev. D51 (1995) 1125

[13] J.C. Collins and D.E. Soper, Nucl. Phys. B194 (1982) 445

[14] B. Guberina, J.H. Kühn, R.D. Pecci and R. Rückl, Nucl. Phys. B174 (1980) 317

[15] G. Altarelli, R.K. Ellis and G. Martinelli, Nucl. Phys. B157 (1979) 461

[16] G. Bodwin and J. Qiu, Phys. Rev. D41 (1990) 2755

[17] J.P.Ma, Melbourne-Preprint, UM–P-95/32, RCHEP–95/11, hep-ph/9504263

[18] G.T. Bodwin, S. Kim and D.K. Sinclair, Nucl. Phys. B(Proc. Suppl.)34 (1994) 434

[19] M. Vänttinen, P. Hoyer, S.J. Brodsky and W.-K. Tang, SLAC-Preprint, SLAC-PUB-6637

**Figure Caption**

Fig.1: The Feynman diagram for the color octet component at the leading order of $\alpha_s$.

Fig.2a: The Feynman diagram for matrix element of the operator $O_8^{Q\bar{Q}}(^3S_1)$ at the leading order of $\alpha_s$ in NRQCD.

Fig.2b: Some Feynman diagrams for a part of the contribution to the matrix element at the order of $\alpha_s$. In this part only the vertex correction and the external leg correction is included.

Fig.2c: Some Feynman diagrams for a part of the contribution to the matrix element at the order of $\alpha_s$. In this part only the contribution induced by a real gluon exchange is included.

Fig.3: The Feynman diagrams for the contribution to the color singlet component at the order of $\alpha_s^2$. They also lead to a part of the contribution to $D_C(Z)$ in Eq.(5.1) for the color octet component.

Fig.4a: The Feynman diagrams for the contribution to $D_B(z)$ in Eq.(5.1).

Fig.4b: The Feynman diagrams for the contribution to $D_C(z)$ in Eq.(5.1).

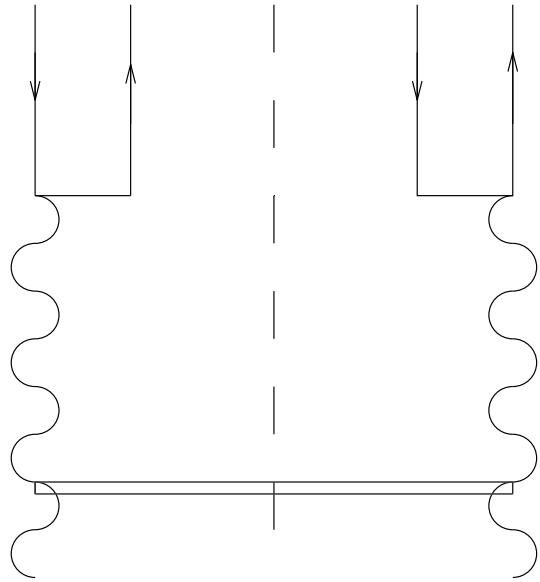

Fig. 1

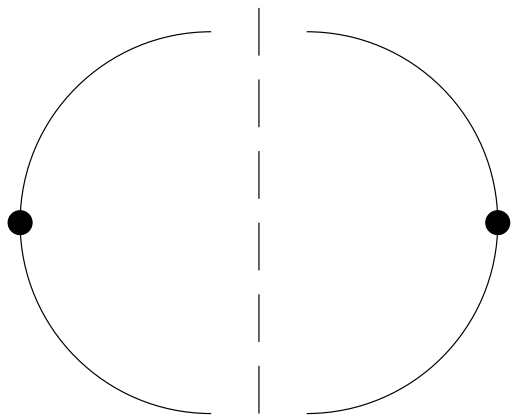

Fig.2a

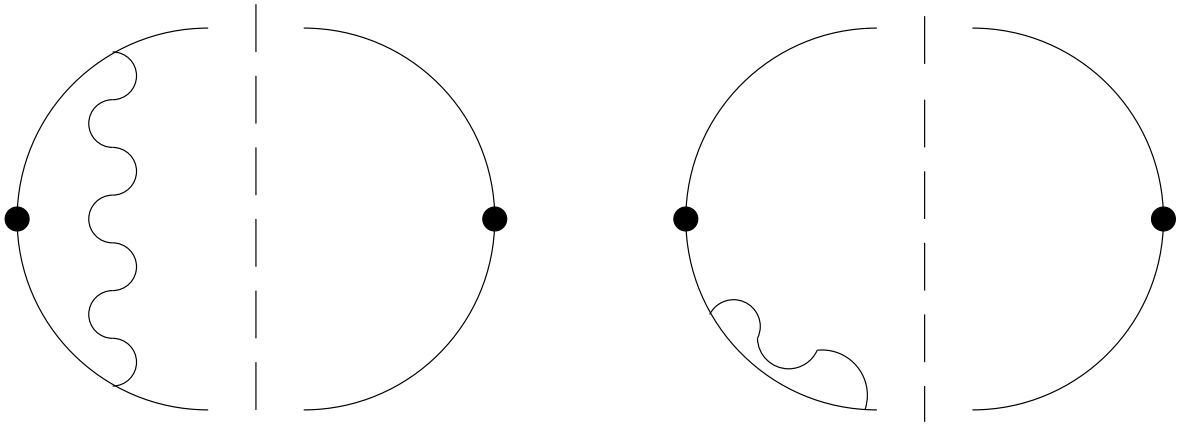

Fig. 2b

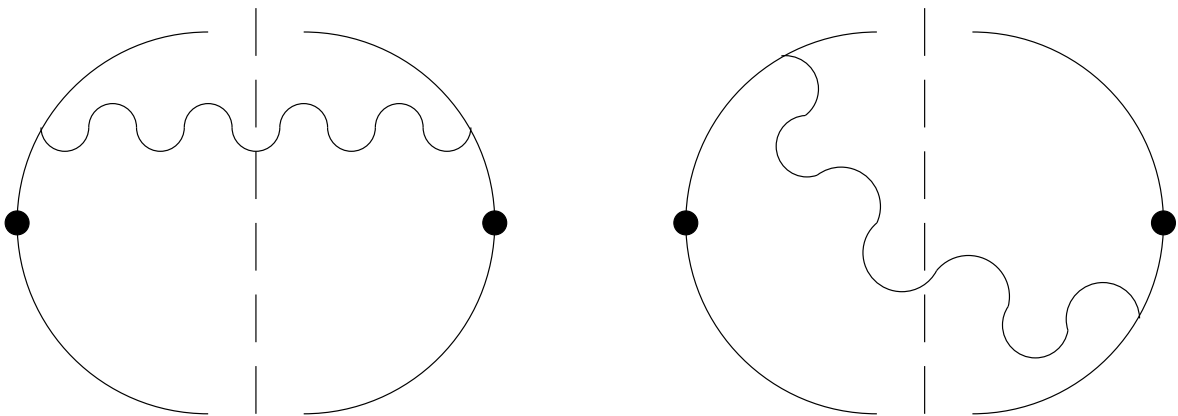

Fig. 2c

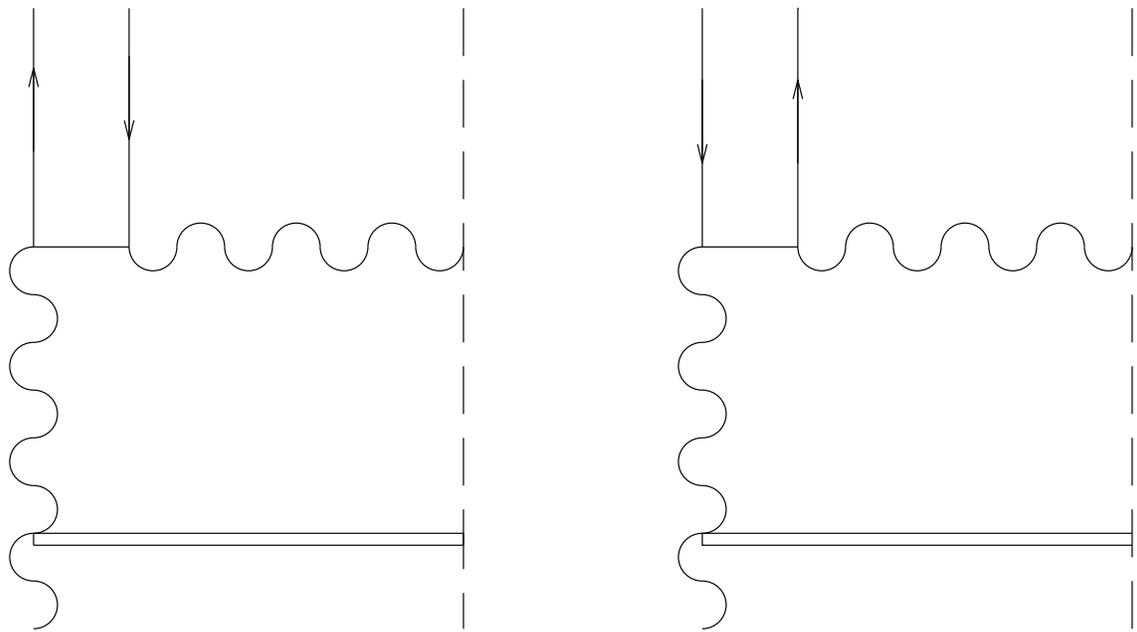

Fig. 3

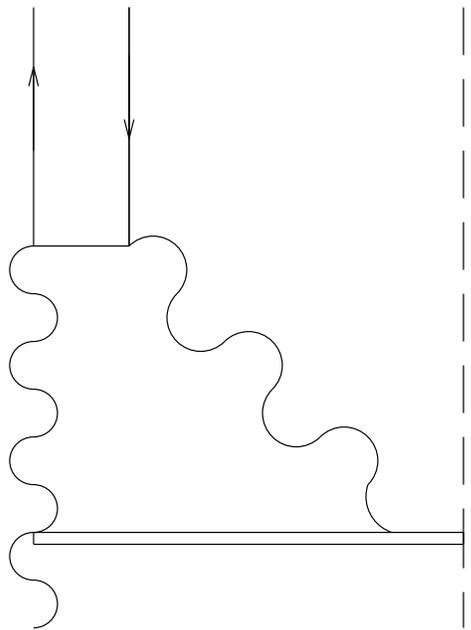 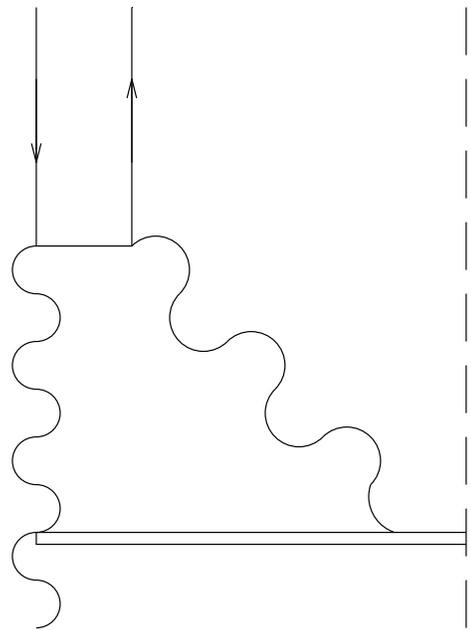

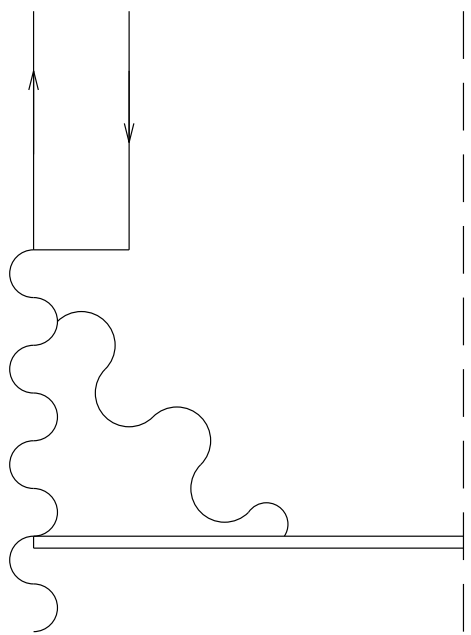 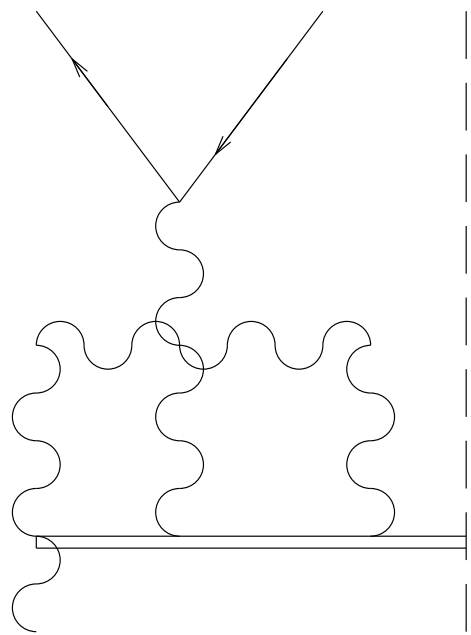

Fig.4a

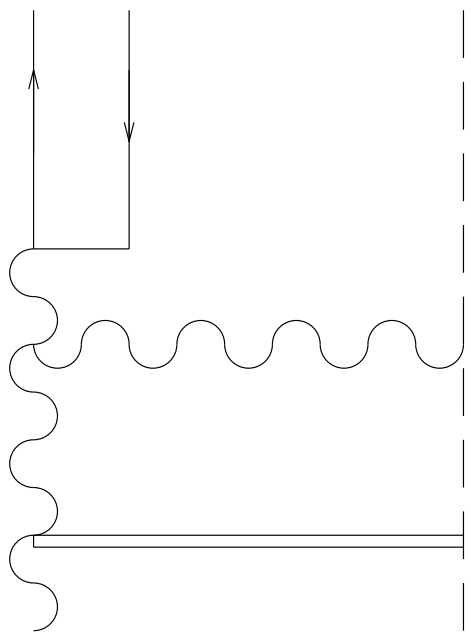
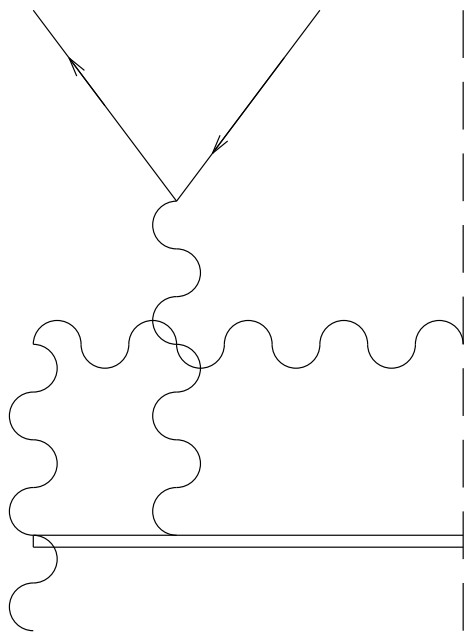

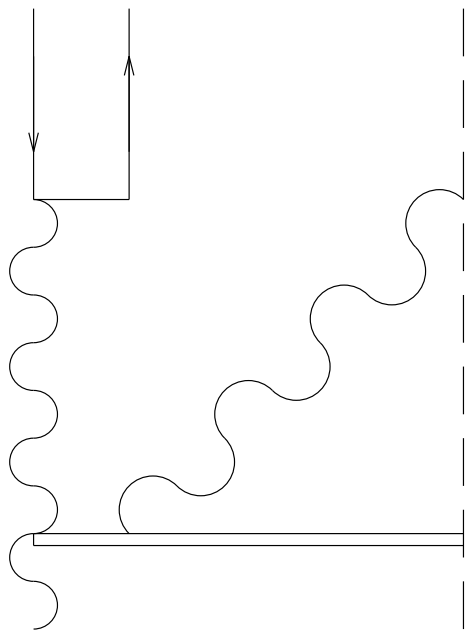

Fig.4b